\documentclass[final,5p,sort,compress]{elsarticle}
\usepackage{tabularx}
\usepackage{amssymb}
\usepackage{hyperref}
\usepackage[table]{xcolor}
\usepackage{tipa}
\usepackage{float}
\usepackage{stfloats}
\usepackage{caption}
\usepackage{booktabs}

\usepackage{etoolbox}
\AtBeginEnvironment{tabularx}{\rowcolors{2}{gray!10}{white}}

\begin{document}

\begin{frontmatter}

\title{PyPackIT: Automated Research Software Engineering for Scientific Python Applications on GitHub}

\author[1]{Armin Ariamajd}
\author[2]{Raquel L\'{o}pez-R\'{i}os de Castro}
\author[1,2]{Andrea Volkamer\corref{cor1}}
\ead{volkamer@cs.uni-saarland.de}

\cortext[cor1]{Corresponding author}

\affiliation[1]{
    organization={Data Driven Drug Design, Faculty of Mathematics and Computer Sciences, Saarland University}, 
    addressline={Campus E2.1},
    postcode={66123},
    city={Saarbrücken},
    country={Germany}
}
\affiliation[2]{
    organization={In silico Toxicology and Structural Bioinformatics, Institute of Physiology, Charité-Universitätsmedizin Berlin},
    addressline={Charitéplatz 1},
    postcode={10117},
    city={Berlin},
    country={Germany}
}

\begin{abstract}
    The increasing importance of Computational Science and Engineering has highlighted the need for high-quality scientific software. However, research software development is often hindered by limited funding, time, staffing, and technical resources. 
    To address these challenges, we introduce PyPackIT, a cloud-based automation tool designed to streamline research software engineering in accordance with FAIR (Findable, Accessible, Interoperable, and Reusable) and Open Science principles.
    PyPackIT is a user-friendly, ready-to-use software that enables scientists to focus on the scientific aspects of their projects while automating repetitive tasks and enforcing best practices throughout the software development life cycle.
    Using modern Continuous software engineering and DevOps methodologies, PyPackIT offers a robust project infrastructure including a build-ready Python package skeleton, a fully operational documentation and test suite, and a control center for dynamic project management and customization.
    PyPackIT integrates seamlessly with GitHub's version control system, issue tracker, and pull-based model to establish a fully-automated software development workflow. Exploiting GitHub Actions, PyPackIT provides a cloud-native Agile development environment using containerization, Configuration-as-Code, and Continuous Integration, Deployment, Testing, Refactoring, and Maintenance pipelines. PyPackIT is an open-source software suite that seamlessly integrates with both new and existing projects via a public GitHub repository template at \url{https://github.com/repodynamics/pypackit}.
\end{abstract}

\begin{graphicalabstract}
    \includegraphics[width=\textwidth]{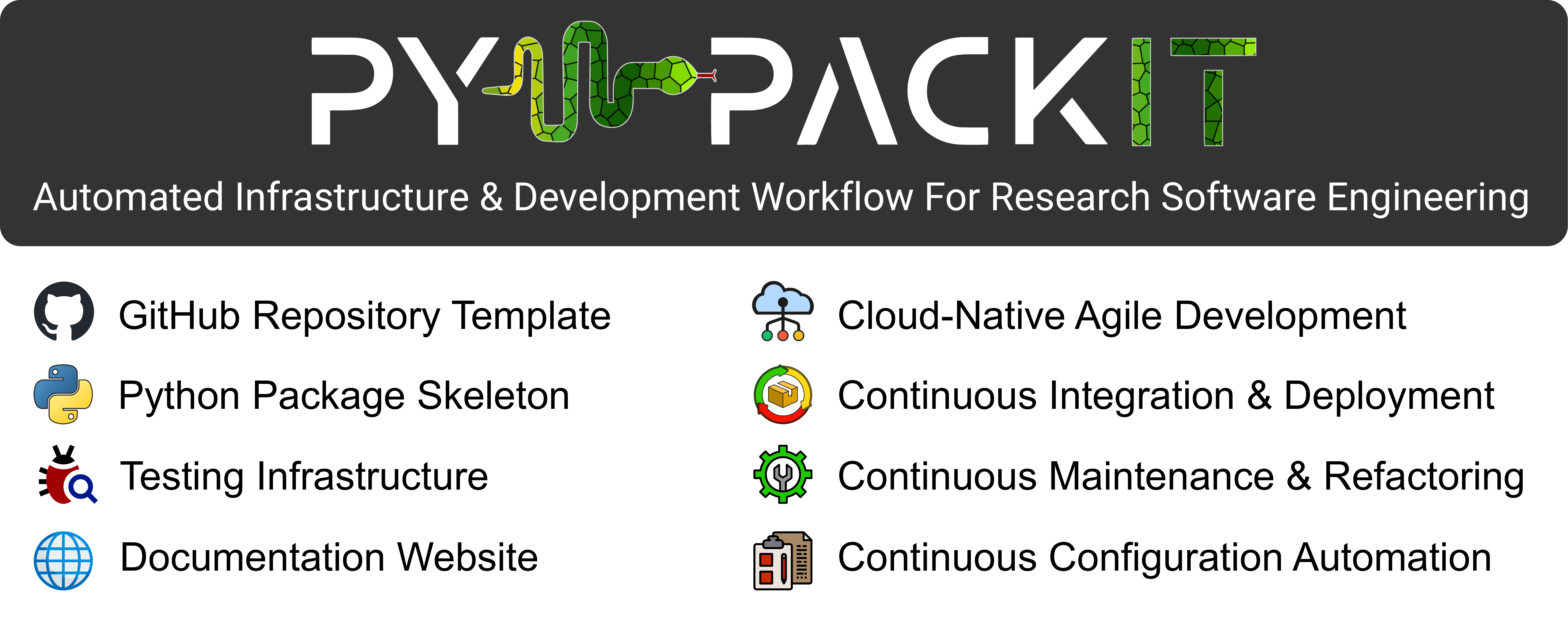}
\end{graphicalabstract}

\begin{highlights}
\begin{minipage}{\textwidth}
\item PyPackIT is a cloud-based automation tool for research software engineering on GitHub 
\item Enables cloud-native Agile development via Continuous software engineering and DevOps
\item Includes Continuous Integration, Deployment, Maintenance, and Configuration pipelines
\item Provides dynamic project skeletons with Python package, test suite, and website
\item PyPackIT is open-source and ready to use through a public GitHub repository template 
\end{minipage}
\end{highlights}


\end{frontmatter}

\sloppy

\section{Introduction}

Computational Science and Engineering (CSE) is a fast-growing discipline that uses computer simulations and numerical methods to gain insights that cannot be achieved through theory and experiments alone \cite{CSE, CSE2, EssenceOfCompSci, Bramley2000}. Software is crucial in CSE for performing simulations as well as generating and analyzing data \cite{RolesOfCodeInCSE, DevelopingSciSoft}. As CSE becomes essential in various scientific fields \cite{ResearchAndEdInCSE}, research software—produced as part of scientific studies—is increasingly featured in publications \cite{UKResearchSoftwareSurvey2014}. Therefore, the validity, reproducibility, and extensibility of these studies strongly depend on the quality of the underlying research software \cite{CompSciDemandsNewParagdim}. 

Research software development is a complex and resource-intensive task, often faced with challenges regarding funding, time, staffing, and technical expertise \cite{HowToSupportOpenSource, ManagingChaos, BetterSoftwareBetterResearch, SoftDevEnvForSciSoft}. Requiring extensive domain knowledge, delegating it to software engineers is a difficult task \cite{SomeChallengesFacingSoftEngsDevSoftForSci, ChallengesFacingSoftEngInSci, WhenEngineersMetScientists}. Thus, in contrast to the software industry where each task is carried out by specialized teams, the entire responsibility of research software development is typically in the hands of small groups of scientists \cite{AnalyzingGitHubRepoOfPapers, HowScientistsReallyUseComputers, NamingThePainInDevSciSoft}. However, due to the growing intricacy of scientific computing and software engineering practices, most scientists have to spend the majority of their time on the scientific aspects of their projects, and are thus self-taught developers with little exposure to modern software engineering methodologies \cite{SoftwareChasm, SciCompGridlock, WheresTheRealBottleneck, SelfPerceptions, SurveySEPracticesInScience2, HowScientistsDevSciSoftExternalRepl}. Therefore, the amount of effort and skills required to produce high-quality research software in accordance with engineering best practices often far exceeds the capabilities of their developers \cite{SoftEngForCompSci, AdoptingSoftEngConceptsInSciResearch, BridgingTheChasm, SurveySEPracticesInScience, HowScientistsDevAndUseSciSoft, UnderstandingHPCCommunity, ProblemsOfEndUserDevs}. This highlights the need for improving research software quality in areas like accessibility, ease of installation and use, documentation, interoperability, extensibility, and maintainability, addressing the so-called \textit{research software crisis} \cite{TroublingTrendsInSciSoftware, ReprodResearchInCompSci, ReproducibleResearchForSciComp, AccessibleReproducibleResearch, SciSoftwareExtensibility, CompSciError, SciSoftwareAccuracy, ExtensibilityAndLibrarization, ShiningLight, TExperiments, WhyJohnnyCantBuild, ImprovingScienceThatUsesCode}.

Acknowledging the importance and challenges of research software development, efforts have been made to improve the status quo. These include the introduction of research software engineering as a new academic discipline \cite{RSEIntro, RSEReportUK, RSEHistory, WhyScienceNeedsMoreRSE, SoftwareSustainabilityInstitute}, and the establishment of various guidelines \cite{FAIR4RS, 4SimpleRecs, 10MetricsForSciSoftware, BestPracticesForSciComp, RecommendOnResearchSoftware, ELIXIRSoftwareManagementPlan, NLeScienceSoftDevGuide, BestPracticesInBioinfSoftware, 10RuleForSoftwareInCompBio, SustainableResearchSoftwareHandOver, QuickGuideToOrgCompBioProjects, EnhancingReproducibility, SciSoftDevIsNotOxymoron, 5RecommendedPracticesForCompSci, 10SimpleRulesOnWritingCleanAndReliableSciSoft, BarelySufficientPracticesInSciComp, GoodEnoughPracticesInSciComp, TuringWay} and workshops \cite{SoftwareCarpentryOriginal, SoftwareCarpentry, SoftEngForSci} to promote software engineering best practices among scientists. However, widespread adoption of such initiatives is often hindered by increased production costs \cite{RSEPillars, RSEinUnis, HowToSupportOpenSource, SoftDevEnvForSciSoft, NamingThePainInDevSciSoft}. For example, integration of engineering best practices can be challenging due to a lack of supporting tools and limited time and knowledge \cite{ConfigManageForLargescaleSciComp}. To improve this situation, we need solutions that are readily accessible and adoptable by all scientists, empowering them to employ research software engineering best practices with ease and minimal overhead \cite{ManagingChaos, SoftEngForCompSci}.

Software engineering involves multiple phases including planning, development, and operations, requiring a well-coordinated workflow using various tools and technologies \cite{CollabSoftEngBookConcepts, StateOfArtInEndUserSoftEng}. By far, the most common problems faced by research software developers are technical issues regarding management, tooling, testing, documentation, deployment, and maintenance of software \cite{NamingThePainInDevSciSoft, ShiningLight, PublishYourCode, BetterSoftwareBetterResearch, SurveySEPracticesInScience, ReprodResearchInCompSci, CaseForOpenCompProg}. Thus, automation tools that streamline such repetitive engineering tasks can significantly accelerate development, improve quality, and lower production costs at the same time \cite{SoftEngForCompSci, BestPracticesForSciComp, AdoptingSoftEngConceptsInSciResearch}. An example proven successful in large-scale scientific initiatives \cite{TrilinosProject} are project skeletons that provide basic infrastructure for software development \cite{ProjectSkeletonsReview}. These are great automation tools for project initiation, but do not support the bulk of repetitive engineering activities that is carried out throughout the development process with increasing complexity and frequency \cite{CollabSoftEngBookConcepts, StateOfArtInEndUserSoftEng, ConfigManageForLargescaleSciComp}. While other general-purpose tools can help streamline individual tasks, comprehensive solutions specifically designed for research software are limited \cite{Bertha, MolSSITemplate, SSCTemplate}.

To fill the current gap, we leveraged our prior experience in scientific software \cite{TeachOpenCADD} to develop PyPackIT—an open-source, ready-to-use, cloud-based automation tool to streamline research software engineering.
PyPackIT supports the entire software development process from initiation to publication and maintenance, enabling scientists to create high-quality and sustainable research software efficiently.

\section{Research Software Engineering Challenges}

In the following, we outline key aspects of modern research software engineering along with a summary of PyPackIT's solution to current challenges (Table \ref{tab:summary}).

\begin{table*}[!ht]
\caption{Overview of key research software engineering aspects, their common advantages and challenges, along with a summary of PyPackIT's solution to these challenges.}
\label{tab:summary}
\begin{tabularx}{\textwidth}{>{\raggedright\arraybackslash}p{2.4cm} >{\raggedright\arraybackslash}X >{\raggedright\arraybackslash}X >{\raggedright\arraybackslash}X}
\toprule
\rowcolor{white} \textbf{Aspect}& \textbf{Advantages} & \textbf{Challenges} & \textbf{PyPackIT's Solution} \\
\midrule

Cloud-native automation & Fast, efficient, and reliable production of high-quality software with reduced cost and risk. & Difficult implementation and maintenance due to configuration, testability, tooling, and security issues. & Ready-to-use development environment and automation workflows with Continuous Integration and Deployment pipelines.\\

Collaborative workflow & Accelerated development and higher code quality through community pull requests and peer reviews. & Effective communication, coordination, documentation, and management. & Pull-based model with automated issue management, organization, task assignment, and documentation.\\

FAIRness & Enhanced findability, accessibility, interoperability, and reusability, ensuring the reproducibility of computational studies. & Efficient mechanisms and tools for packaging, distribution and indexing. & Automated licensing, build, containerization, and deployment to multiple indexing repositories with comprehensive metadata and identifiers.\\

Quality assurance & Established correctness and functionality with improved maintainability, performance, and security. & Consistent and effective quality assurance requires automated cloud-native solutions. & Ready-to-use test suite and testing infrastructure integrated into Continuous Integration, Refactoring, and Testing pipelines.\\

Documentation & Instructions to install and use software, clearly documenting its capabilities, limitations, and changes over time. & Considerable time, effort, and skills required to develop, deploy, and continuously update documentation. & Fully designed and customizable documentation website with automatic content generation, deployment, and maintenance.\\

Version control & Parallel and isolated development with efficient code management, annotation, change tracking, and backup. & An established strategy tailored to research software needs and consistently applied throughout its life cycle. & Automated version control workflows with specialized branching model and versioning scheme.\\

Configuration & Required for providing project-specific metadata and settings to various tools and services used in the project. & Static tool-specific files and manual settings complicate configuration maintenance and customization. & Single interface to unify settings and establish Continuous Configuration Automation in the entire project.\\

Maintenance & Long-term usability and sustainability, enabling reproducibility of studies and preventing reimplementation. & Consistent effort required to keep the software operational and improve functionalities. & Development containers with Continuous Maintenance, Refactoring, and Testing pipelines to minimize software entropy and technical debt.\\
\bottomrule
\end{tabularx}
\end{table*}

\subsection{Cloud-Native Automation}

Research software often faces evolving requirements so that determining the exact specifications of the end product is usually not possible in advance \cite{SoftDevEnvForSciSoft, ProblemsOfEndUserDevs}. Consequently, as traditional development methodologies may not accommodate effectively for research software \cite{DevelopingSciSoft, SoftEngForCompSci, DealingWithRiskInSciSoft, BalancingAgilityAndDiscipline}, cloud-native practices such as Agile development, Continuous software engineering, and DevOps are recommended \cite{AdoptingSoftEngConceptsInSciResearch, LitRevAgileInSciSoftDev, BestPracticesForSciComp}.
Agile development \cite{AgileSoftDev}, based on iterative enhancement and short feedback cycles, aligns well with the experimental nature of research software \cite{WhenEngineersMetScientists, SoftDevEnvForSciSoft, SurveySEPracticesInScience}, effectively managing high rates of change, complexity, and risk \cite{AgileSoftDevMethodAndPractices, AgileSoftDevEcosystems}. In tandem, Continuous software engineering practices including Continuous Integration (CI), Continuous Delivery (CDE), and Continuous Deployment (CD)—collectively called CI/CD—enhance the efficiency, scalability, and reliability of Agile development through automation \cite{ContSoftEngineering, ContinuousSoftEng, CICDSystematicReview}. They provide numerous benefits, including decreased errors, more efficient bug discovery, and a high level of control over applied changes, allowing projects to produce higher quality software more rapidly, efficiently, and reliably \cite{UsageCostsAndBenefitsOfCI, CDatFacebook, EffectsOfCIOnSoftDev, QualityAndProductivityCI, ImpactOfCI, ExpBenefitsOfCI, UncoveringBenefitsAndChallengesOfCI, ModelingCI, ExtremeProgExplained, StairwayToHeaven, CDReliableSoftReleaseBook, CDatFBandOANDA, CDHugeBenefits, StudyImpactAdoptCIOnPR}. DevOps extends these practices by automating the integration of Development (Dev) and Operations (Ops) phases, further benefiting research software projects \cite{WhatIsDevOps, DevOpsInSciSysDev, ResearchOps}.

While cloud-native methodologies are considered crucial \cite{EffectsOfCIOnSoftDev, CICDSystematicReview, AnalysisOfTrendsInProductivity} and are well-established in industry \cite{EmpEvAgile, AgileAdoptionSurvey, Top10AdagesInCD, SynthCDPractices} and some large research institutions \cite{IntroducingAgileInBioInf, AgileInBioMedSoftDev, UsingAgileToDevCompBioSoft, ExploringXPForSciRes}, their adoption in academia presents opportunities for growth \cite{SurveySEPracticesInScience2, SelfPerceptions, AdoptingSoftEngConceptsInSciResearch, DevelopingSciSoft}. As implementing Continuous pipelines is complex and ready-to-use solutions are scarce \cite{ModelingCI, CICDSystematicReview, UncoveringBenefitsAndChallengesOfCI, StairwayToHeaven, CDHugeBenefits, HowDoSoftDevsUseGHA, DevPerceptionOfGHA, EffectsOfCIOnSoftDev}, many open-source projects are in need of solutions to integrate cloud-native practices \cite{CITheater, AutoSecurityAssessOfGHAWorkflows, AmbushFromAllSides, OnOutdatednessOfWorkflowsInGHA}. \href{https://github.com}{GitHub} offers public repositories free integration with \href{https://github.com/features/actions}{GitHub Actions} (GHA)—an event-driven cloud computing platform for execution of automated software development \href{https://docs.github.com/en/actions/using-workflows/about-workflows}{workflows}, modularizable into reusable components called \href{https://docs.github.com/en/actions/creating-actions/about-custom-actions}{actions} \cite{GitHubDevWorkflowAutoEcoBook, HandsOnGHA}. Currently the most popular CI/CD service on GitHub \cite{RiseAndFallOfCIinGH, OnUsageAndMigrationOfCITools, OnUseOfGHA}, GHA grants workflows full control of all repository components, enabling automation well beyond conventional CI/CD practices \cite{DevPerceptionOfGHA, GitHubDevWorkflowAutoEcoBook}. However, implementing GHA workflows and actions involves challenges related to tooling, configuration, testability, debugging, maintenance, and security \cite{HowDoSoftDevsUseGHA, EvolutionOfGHAWorkflows, OnOutdatednessOfWorkflowsInGHA, AutoSecurityAssessOfGHAWorkflows}. Consequently, most projects do not make use of these advanced features that can greatly improve the software development process \cite{OnUseOfGHA, LetsSuperchargeWorkflows}.

\subsection{Collaborative Workflow}

Software development is increasingly collaborative and distributed, requiring a robust workflow for effective communication and coordination \cite{ScaleAndEvolOfCoordNeeds, GlobalSoftEng, InfluenceOfSocialAndTechnicalFactors, UnderstandingCommunitySmells, GlobalSoftDevChallenges, ConfigManageForLargescaleSciComp, CollabSoftEngBookConcepts, StateOfArtInEndUserSoftEng}. Many research software projects suffer from management and maintenance issues due to lacking workflows \cite{ProblemsOfEndUserDevs, DevelopingSciSoft, SurveySEPracticesInScience, NamingThePainInDevSciSoft, ConfigManageForLargescaleSciComp}. Cloud-based social coding platforms (SCPs) address these challenges by providing essential software engineering tools in a transparent mutual environment \cite{OpenSourceSoftHostingPlatforms, CharacterizingProjEvolOnSocialCodingPlat, SocialCodingInGitHub}, including distributed version control systems (VCSs) like \href{https://git-scm.com/}{Git} \cite{VCSReview}. GitHub, currently the largest SCP \cite{GitHubOctoverse2024}, is especially recommended for research projects \cite{10RuleForSoftwareInCompBio, BestPracticesForSciComp, 10SimpleRulesGitAndGitHub} as it provides special features like software citation and free upgrades for academic use \cite{GitHubForScience}. GitHub's pull-based development model offers an effective solution by enabling community contributions through issue tickets and pull requests (PRs), while maintainers review and integrate changes \cite{ExplorStudyPullBased, WorkPractPullBased, CharacterizingProjEvolOnSocialCodingPlat}. 
This accelerates development and enhances code quality through reviews \cite{5RecommendedPracticesForCompSci, 10MetricsForSciSoftware, BestPracticesForSciComp}, but also requires careful management in terms of task assignment, issue–commit linkage, and documentation \cite{WhatMakesCompSoftSuccessful, DealingWithRiskInSciSoft, 5RecommendedPracticesForCompSci, CharacterizingProjEvolOnSocialCodingPlat, SciSoftDevIsNotOxymoron, 4SimpleRecs, SustainableResearchSoftwareHandOver, ConfigManageForLargescaleSciComp}.

Issue tracking systems (ITSs) like GitHub Issues (GHI) help document and organize tasks, but need significant setup to function effectively \cite{SurveySEPracticesInScience, DLRSoftEngGuidelines, ELIXIRSoftwareManagementPlan, EmpAnalysisOfIssueTemplatesOnGitHub}. By default, GHI only offers a single option for opening unstructured issue tickets, which can lead to problems such as missing crucial information that complicate issue triage \cite{EmpAnalysisOfIssueTemplatesOnGitHub}. To facilitate issue management, GitHub offers labeling features to help categorize and prioritize tickets \cite{GiLaGitHubLabelAnalyzer, ExploringCharacIssueRelatedGitHub}. However, management tasks must be done manually, which is time-consuming and prone to errors \cite{WhereIsTheRoadForIssueReports, GotIssues, ExploringTheUseOfLabels, FillingTheGapsOfDevLogs, MissingLinksBugsAndBugFix}, motivating the development of machine-learning tools for automatic ticket classification \cite{PredictingIssueTypesOnGitHub, ImpactOfDataQualityForAutomaticIssueClassification} and issue–commit link recovery \cite{IssueCommitLink-DeepLink, FRLink}. Recently, GitHub introduced \href{https://github.blog/changelog/2021-06-23-issues-forms-beta-for-public-repositories/}{issue forms}; customizable web forms that enable the collection of machine-readable user inputs \cite{EmpAnalysisOfIssueTemplatesOnGitHub, FirstLookAtBugReportTempOnGitHub, UnderstandingIssueTemplateOnGitHub}. While they can be used in conjunction with GHA to automate issue management tasks without the need for machine-learning tools, these capabilities are often not exploited due to the initial implementation barrier.

\subsection{FAIRness}

Research software is vital for computational studies but often lacks Findability, Accessibility, Interoperability, and Reusability—key aspects of the FAIR principles \cite{FAIR4RS}—impacting reproducibility and reuse of scientific studies \cite{AccessibleReproducibleResearch, ShiningLight, CaseForOpenCompProg, SciSoftwareAccuracy, SurveySEPracticesInScience}. \textbf{Findability} requires that research software is searchable by its functionalities and attributes, necessitating its permanent distribution to related public indexing repositories along with comprehensive metadata and unique identifiers like DOIs \cite{10MetricsForSciSoftware, WhatMakesCompSoftSuccessful, 10SimpleRulesForOpenDevOfSciSoft, 4SimpleRecs, BarelySufficientPracticesInSciComp, ELIXIRSoftwareManagementPlan}. \textbf{Accessibility} involves adopting an open-source model under a permissive license \cite{BusinessOfOpenSource}—ideally from the start \cite{BetterSoftwareBetterResearch, PublishYourCode, POVHowOpenSciHelps}—to enable transparent peer reviews, facilitate progress tracking, and promote trust, adoption, and collaboration \cite{SharingDetailedResData, CaseForOpenCompProg, 10SimpleRulesForOpenDevOfSciSoft}. For \textbf{interoperability}, a key factor is using a well-suited and popular programming language in the target community \cite{RolesOfCodeInCSE, SoftDevEnvForSciSoft}. Python is now the leading language for research software development \cite{SurveySEPracticesInScience2, AnalyzingGitHubRepoOfPapers, DevOpsInSciSysDev}, recommended due to its simplicity, versatility, and ability to quickly implement complex tasks that are hard to address in low-level languages \cite{PythonBatteriesIncluded, PythonForSciComp, PythonForSciAndEng, PythonJupyterEcosystem, SciCompWithPythonOnHPC, PythonEcosystemSciComp, WhatMakesPythonFirstChoice}. Python's extensive ecosystem offers performance-optimized libraries for various scientific applications \cite{NumPy, SciPy, pandas, PyTorch, Top5MLLibPython, ScikitLearn, scikitImage, Matplotlib, Mayavi, IPython, Jupyter, Jupyter2, DaskAndNumba, DaskApplications, Cython, Numba, Pythran, PyCUDA, Astropy, SunPy, Pangeo, MDAnalysis, Biopython, NIPY}, with features for parallel distributed computing \cite{SciCompWithPythonOnHPC, ParallelDistCompUsingPython, ScientistsGuideToCloudComputing, DemystPythonPackageWithCondaEnvMod, PythonAcceleratorsForHPC, DistWorkflowsWithJupyter, InteractiveSupercomputingWithJupyter} that even allow high-performance computing (HPC) communities such as CERN \cite{IntroducingPythonAtCERN, PythonAtCERN} and NASA \cite{PythonAtNASA} to use it for key scientific achievements \cite{PythonScientificSuccessStories, GravWaveDiscovery, BlackHoleImage}. Lastly, \textbf{reusability} is enabled by employing DRY (Don't Repeat Yourself) principles and modularizing code into applications with clear programming and user interfaces \cite{FAIR4RS, 5RecommendedPracticesForCompSci, BestPracticesForSciComp, RolesOfCodeInCSE}. Applications must then be packaged into as many distribution formats as possible, to ensure compatibility with different hardware and software environments. This can also greatly simplify the setup process for users \cite{10RuleForSoftwareInCompBio, ELIXIRSoftwareManagementPlan, WhyJohnnyCantBuild}, which is a common problem in research software \cite{NamingThePainInDevSciSoft, CompSciError}. 
To ensure the reproducibility of computational studies, consistent execution and predictable outcomes must be guaranteed regardless of the runtime environment. This is achieved through containerization—a cloud-native approach using technologies like Docker to encapsulate applications and all their dependencies into isolated, portable images \cite{AdoptingSoftEngConceptsInSciResearch, 10RuleForSoftwareInCompBio, ELIXIRSoftwareManagementPlan, Docker}.

Despite its importance, research software is infrequently published \cite{AnalyzingGitHubRepoOfPapers, BridgingTheChasm, PublishYourCode, CompSciError}, leading to controversies and paper retractions \cite{InfluentialPandemicSimulation, RetractionCOVID}, and forcing the reimplementation of computational workflows from scratch \cite{ProblemsOfEndUserDevs, BetterSoftwareBetterResearch, SurveySEPracticesInScience2, SoftEngForCompSci}. Thus, there is a growing call for an open research culture to enhance transparency and reproducibility \cite{PromotingOpenResearch, ReprodResearchInCompSci, EnhancingReproducibility, TroublingTrendsInSciSoftware}, and many journals now mandate source code submissions for peer-review and public access \cite{RealSoftwareCrisis, DoesYourCodeStandUp, TowardReproducibleCompResearch, MakingDataMaximallyAvailable, JournalOfBioStatPolicy}. This highlights the need for efficient tools and mechanisms for licensing, packaging, containerization, distribution, indexing, and maintenance—key challenges in publishing FAIR research software \cite{CaseForOpenCompProg, SurveySEPracticesInScience, ReprodResearchInCompSci, BarelySufficientPracticesInSciComp, BetterSoftwareBetterResearch, PublishYourCode}.

\subsection{Quality Assurance and Testing}

Given the integral role of computational studies in solving critical real-life problems, ensuring the correctness of scientific software is of utmost importance, as errors can lead to inaccurate conclusions \cite{CompSciError, BestPracticesForSciComp, 5RecommendedPracticesForCompSci, SciSoftDevIsNotOxymoron, SurveySEPracticesInScience}. Due to the high complexity of research software, test-driven development is essential to prevent errors \cite{SoftDevEnvForSciSoft, EmpStudyDesignInHPC}. It involves early and frequent unit and regression testing to validate new code components and ensure existing features remain functional after changes \cite{10SimpleRulesOnWritingCleanAndReliableSciSoft, SurveySEPracticesInScience, BarelySufficientPracticesInSciComp, 10SimpleRulesOnWritingCleanAndReliableSciSoft, BestPracticesForSciComp}. To establish testing effectiveness, coverage metrics must be frequently monitored to identify untested components \cite{DLRSoftEngGuidelines, 10SimpleRulesOnWritingCleanAndReliableSciSoft}. Users should also be able to run tests locally to verify software functionality and performance on their machines \cite{ELIXIRSoftwareManagementPlan, DLRSoftEngGuidelines}, necessitating the tests to be packaged and distributed along with the software \cite{BestPracticesInBioinfSoftware, BarelySufficientPracticesInSciComp, 10MetricsForSciSoftware}. Other crucial quality assurance routines include static code analysis such as linting and type checking to identify issues undetected by tests, and code formatting to improve readability and maintain a consistent style \cite{DLRSoftEngGuidelines, BestPracticesForSciComp, SurveySEPracticesInScience, 10SimpleRulesOnWritingCleanAndReliableSciSoft, NLeScienceSoftDevGuide}. Automating these practices in CI/CD pipelines is essential for effective code quality assurance \cite{BestPracticesForSciComp, 10MetricsForSciSoftware, 10SimpleRulesOnWritingCleanAndReliableSciSoft, SurveySEPracticesInScience}. This is however a challenging task \cite{StairwayToHeaven}, resulting in the prevalence of slow and ineffective testing methods that can compromise scientific findings \cite{TestingResearchSoftwareSurvey, SoftEngForCompSci, SurveySEPracticesInScience2, SoftwareChasm, SurveySEPracticesInScience, CompSciError, ApproxTowerInCompSci} and lead to errors, paper retractions, and corrections in research literature \cite{NightmareRetraction, RetractionChang, RetractionChang2, RetractionMa, RetractionJAmCollCardiol, RetractionMeasuresOfCladeConfidence, RetractionsEffectOfAProgram, CorrectionHypertension, CommentOnError, CommentOnError2, CommentOnError3, CommentOnError4, CommentOnError5, ClusterFailureFMRI}. 

\subsection{Documentation}

Documentation is a key factor in software quality and success, ensuring users understand how to install, use, and exploit the software's capabilities while recognizing its limitations \cite{10SimpleRulesForOpenDevOfSciSoft, BestPracticesForSciComp, GoodEnoughPracticesInSciComp, WhatMakesCompSoftSuccessful, SciSoftDevIsNotOxymoron, NamingThePainInDevSciSoft, CompSciError, BarelySufficientPracticesInSciComp}. This is especially important for research software, which suffers from knowledge loss due to high developer turnover \cite{RecommendOnResearchSoftware, EmpStudyDesignInHPC, SoftwareSustainabilityInstitute}. Moreover, as software evolves, documenting and publishing changelogs with each release is crucial to allow existing users assess the update impact and help new contributors understand the software's progression \cite{ELIXIRSoftwareManagementPlan, GoodEnoughPracticesInSciComp, SustainableResearchSoftwareHandOver}. Since community building is crucial for success \cite{WhatMakesCompSoftSuccessful, HowToSupportOpenSource}, project documentation should also include contribution guidelines, governance models, and codes of conduct \cite{SurveySEPracticesInScience, BestPracticesForSciComp, BestPracticesInBioinfSoftware, SustainableResearchSoftwareHandOver, 4SimpleRecs, ELIXIRSoftwareManagementPlan, DLRSoftEngGuidelines, NLeScienceSoftDevGuide}. However, high-quality documentation requires time, effort, and skills, including web development knowledge to create user-friendly websites that stay up to date with the latest project developments \cite{SurveySEPracticesInScience, WhatMakesCompSoftSuccessful}. Although tools exist to aid documentation \cite{TenSimpleRulesForDocumentingSciSoft, WhatMakesCompSoftSuccessful, BestPracticesForSciComp}, developers must still invest time in setting them up. Another important issue is the lack of automation, requiring manual documentation of a large amount of instructions and specifications. Consequently, research software is often not well-documented \cite{SoftEngForCompSci, ProblemsOfEndUserDevs, AnalyzingGitHubRepoOfPapers, DealingWithRiskInSciSoft}, creating barriers to use and leading to software misuse and faulty scientific results \cite{HowScientistsReallyUseComputers, HowScientistsDevSciSoftExternalRepl, CompSciError}.

\subsection{Version Control}

Version control practices such as branching and tagging are vital yet challenging tasks in software development \cite{10MetricsForSciSoftware, ELIXIRSoftwareManagementPlan, EffectOfBranchingStrategies, BranchUseInPractice}. Branching provides isolation for development and testing of individual changes, which must then be merged back into the project's mainline with information-rich commit messages to maintain a clear history. Moreover, tags allow to annotate specific states of the code with version numbers to clearly communicate and reference changes \cite{ImportanceOfBranchingModels, CICDSystematicReview}. While established versioning schemes like \href{https://semver.org/}{Semantic Versioning} and branching models like git-flow and trunk-based development exist \cite{TrunkBasedDev, GitFlow, GitHubFlow, GitLabFlow}, consistently enforcing them still requires an automated workflow. Additionally, general-purpose strategies may require adjustments to fully align with the evolving nature of research software, which often begins as a prototype and undergoes significant changes \cite{UnderstandingHPCCommunity}. Therefore, a suitable model should support simultaneous development and long-term maintenance of multiple versions, to facilitate rapid evolution while ensuring the reproducibility of earlier studies \cite{ConfigManageForLargescaleSciComp}. 

\subsection{Configuration Management}

Software projects rely on various tools and services throughout the development life cycle, each requiring configuration via specific files or user interfaces. This can lead to several maintenance challenges \cite{BestPracticesForSciComp, DevOpsInSciSysDev}: Configuration files are often static and require manual updates to reflect changes. Otherwise, they can quickly fall out of sync with the project's current state, leading to conflicts and inconsistencies. Configurations via interactive user interfaces further complicate the tracking and replication of settings, as they must be manually recorded and applied. Moreover, tool-specific formats and requirements result in data redundancy, as many settings are shared. 
DevOps practices such as Continuous Configuration Automation (CCA), Configuration-as-Code (CaC), and Infrastructure-as-Code (IaC) were developed to tackle these issues, enabling dynamic configuration management of hardware and software infrastructures through machine-readable definition files \cite{InfrastructureAsCode}. While these practices are more prevalent in network management applications, they can greatly benefit software development projects as well. However, due to limited tool availability, most projects still rely on a combination of different configuration files and manual settings, causing inefficiencies, misconfigurations, and poor reproducibility.

\subsection{Maintenance}

Modern research software can remain useful and operational for decades, making its maintenance crucial to reflect ongoing scientific advances \cite{SoftwareSustainabilityInstitute, SoftEngForCompSci, SoftDevEnvForSciSoft}. Sustaining high-quality software requires continuous feedback from the community and active maintenance to address issues, improve functionality, and add new features \cite{BarelySufficientPracticesInSciComp, SoftwareSustainabilityInstitute}. Maintaining dependencies is equally important \cite{EmpComparisonOfDepNetEvolution, FortyYearsOfSoftwareReuse}, as software must remain compatible with diverse environments and future dependency versions. However, many projects overlook outdated dependencies, leading to incompatibilities and bugs \cite{DoDevsUpdateDeps, MeasuringDepFreshness, ThouShaltNotDepend, OnImpactOfSecVulnInDepNet}. Challenges such as short-term funding \cite{ManagingChaos, BetterSoftwareBetterResearch}, small team sizes \cite{SoftEngForCompSci, HowScientistsReallyUseComputers}, and high developer turnover rates \cite{RecommendOnResearchSoftware, EmpStudyDesignInHPC} further hinder maintenance, exacerbated by technical debt from neglected software engineering best practices \cite{BetterSoftwareBetterResearch, ProblemsOfEndUserDevs, SoftEngForCompSci, ManagingTechnicalDebt, 10SimpleRulesForOpenDevOfSciSoft, SoftDesignForEmpoweringSci, ManagingChaos, SoftwareSustainabilityInstitute}. Consequently, the extra effort required for maintenance is a major barrier to publicly releasing research software \cite{BetterSoftwareBetterResearch, PublishYourCode}, often leaving it as an unsustainable prototype \cite{SustainableResearchSoftwareHandOver, 10RuleForSoftwareInCompBio, PublishYourCode}. To prevent such issues, Continuous Maintenance \cite{ContinuousMaintenance}, Refactoring \cite{ContRefact}, and Testing \cite{ContinuousSoftEng} pipelines are required to periodically update dependencies and development tools, and automatically maintain the health of the software and its development environment from project start \cite{SoftEngForCompSci}. Furthermore, providing a ready-to-use development environment can significantly reduce the barrier to entry for new contributors and maintainers, fostering the long-term sustainability of research software.

\section{Software Overview}
\label{section-overview}

PyPackIT is a ready-to-use automation tool, fully configured in accordance with research software requirements. It is an open-source and cloud-based software suite, hosted on GitHub at \url{https://github.com/repodynamics}. PyPackIT comprises several GitHub Actions (GHA) actions and Python applications, which can be installed in GitHub repositories to perform automated tasks on the GHA cloud framework. To facilitate installation, PyPackIT includes a \href{https://docs.github.com/en/repositories/creating-and-managing-repositories/creating-a-repository-from-a-template}{repository template} at \url{https://github.com/repodynamics/pypackit}. There, by clicking the \href{https://github.com/new?template_name=PyPackIT&template_owner=RepoDynamics}{\texttt{Use this template}} button, users can create a new GitHub repository that automatically contains all the necessary files to run PyPackIT, such as GHA workflow definition files. Installation in existing repositories is supported as well, simply requiring users to add these files. Detailed installation guides are provided on PyPackIT's documentation website at \url{https://pypackit.repodynamics.com}, along with comprehensive user manuals, tutorials, and technical information.

Upon installation, users only need to invest a few minutes filling project-specific information in the provided configuration files. PyPackIT then takes over, automatically setting up the repository and generating a complete infrastructure for the project, including a build-ready python package, fully automated test suite and documentation website, license and citation files, community health files like dynamic repository READMEs, and configuration files for various development tools and services. An initial release of the software along with its test suite and documentation website can be immediately deployed online, registering the project in various indexing repositories. All metadata and settings are readily customizable via PyPackIT's configuration files, which form a unified control center enabling Continuous Configuration Automation (CCA) for the entire project infrastructure and development environment throughout the software life cycle.

\begin{figure*}[t]
    \centering
    \includegraphics[width=1\linewidth]{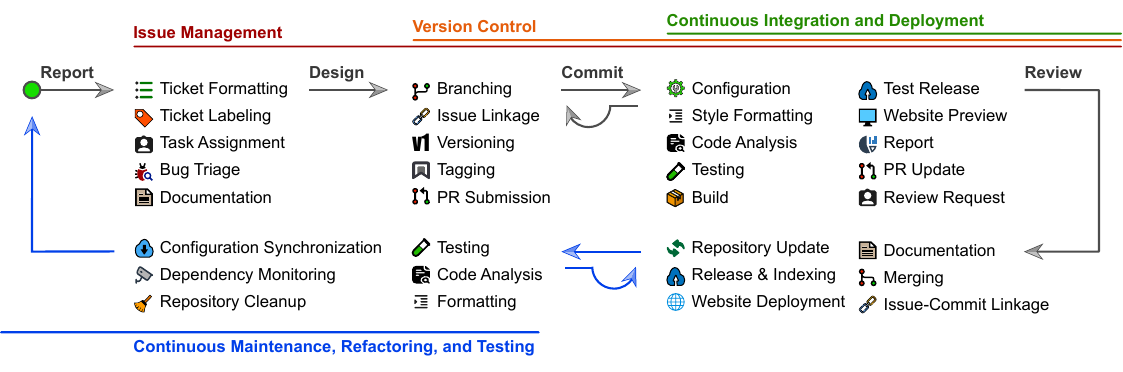}
    \caption{PyPackIT's software development workflow. Labeled arrows represent manual tasks performed by users: Report, Design, Commit, and Review. All other activities are automated, which fall into four main categories spanning different stages of the software development life cycle: Issue Management, Version Control, Continuous Integration and Deployment, and Continuous Maintenance, Refactoring, and Testing.}
    \label{fig:workflow}
\end{figure*}

After installation, PyPackIT activates automatically in response to a range of GitHub repository events, such as issue ticket submissions, pushing commits, and pull request (PR) reviews. It then analyzes the triggering event and the current state of the repository to determine the appropriate response, establishing a cloud-native Agile software development process with an automated pull-based workflow. A schematic overview of PyPackIT's software development workflow is illustrated in Figure \ref{fig:workflow}. This workflow leaves users with only the following four manual tasks throughout the software life cycle:

\begin{itemize}
    \item \textbf{Report}: Each task in the project starts by submitting a ticket to its issue tracking system (ITS). PyPackIT facilitates issue reporting by automatically configuring and maintaining GitHub Issues (GHI) to provide users with specialized submission forms for various issue types. Following a ticket submission, PyPackIT automatically performs issue management tasks, reducing the triage process to a simple decision on whether to implement the task.
    \item \textbf{Design}: Once an issue ticket is approved for implementation, users simply need to complete a brief development document by filling out a form attached to the ticket. PyPackIT's automatic version control then activates, creating a ready-to-use branch for implementation along with dynamically maintained PRs, changelogs, and draft releases for documenting changes and tracking progress.
    \item \textbf{Commit}: With PyPackIT's comprehensive infrastructure, implementing changes is streamlined to writing scientific code, docstrings, and test cases in the provided skeleton files, or modifying project configurations in the control center, depending on the task. Each commit triggers PyPackIT's Continuous Integration and Deployment (CI/CD) pipelines, which automatically integrate the new changes into the project. These pipelines perform various quality assurance and testing tasks while generating version tags, reports, build artifacts, Docker images, and developmental releases for review.
    \item \textbf{Review}: When the implementation is finished, PyPackIT automatically sends review requests to designated maintainers. Once approved, PyPackIT's CI/CD pipelines merge the changes into the project's mainline, updating all affected project components and generating detailed changelogs that document the entire development process. If changes correspond to software's public interfaces, a new version is generated along with release notes and various artifacts, which can be automatically published to multiple indexing repositories (Table \ref{table:indexing-repos}). 
\end{itemize}

\begin{table*}[t]
\caption{Indexing repositories supported by PyPackIT. Deployment is automated by PyPackIT's CD pipeline, requiring only a one-time setup. This can be done by configuring \href{https://docs.pypi.org/trusted-publishers/}{Trusted Publishing} for the chosen indexing repository or by generating an access token and adding it as a secret to the GitHub repository.}
\label{table:indexing-repos}
\begin{tabularx}{\textwidth}{>{\raggedright\arraybackslash}p{2.6cm} X >{\raggedright\arraybackslash}p{3.1cm}}
\toprule
\rowcolor{white}\textbf{Repository} & \textbf{Description} & \textbf{Requirement}\\
\midrule

\href{https://pypi.org/}{PyPI} & Python Package Index; Python's official software repository used by the \href{https://pip.pypa.io/}{Pip} package manager. & Trusted Publishing \\

\href{https://anaconda.org/}{Anaconda} & A language-agnostic repository popular in the scientific community and used by the \href{https://docs.conda.io}{Conda} package manager. & Token \\

Docker & Any Docker registry such as \href{https://hub.docker.com/}{Docker Hub} or \href{https://github.blog/news-insights/product-news/introducing-github-container-registry/}{GitHub Container Registry} (GHCR). & Token (not required for GHCR)\\

\href{https://zenodo.org/}{Zenodo} & A repository capable of minting persistent DOIs for immutable depositions. & Token\\

\href{https://docs.github.com/en/repositories/releasing-projects-on-github/about-releases}{GitHub Releases} & GitHub's repository for uploading build artifacts and release notes. & No requirements\\

\href{https://test.pypi.org/}{TestPyPI} & A separate instance of PyPI for testing purposes. & Trusted Publishing\\

\href{https://sandbox.zenodo.org/}{Zenodo Sandbox} & A separate instance of Zenodo for testing purposes. & Token\\
\bottomrule
\end{tabularx}
\end{table*}

Moreover, PyPackIT also operates on a scheduled basis, executing Continuous Maintenance, Refactoring, and Testing pipelines throughout the project's lifespan. These perform various maintenance and monitoring tasks on the repository and previously released software versions, ensuring the long-term health of scientific applications and their development environment. When actions are needed, PyPackIT can automatically submit issue tickets for manual implementation, PRs for automatic fixes that require review, or it can directly update the project as necessary. 
To further support developers during the implementation phase, PyPackIT encapsulates the project's development environment into \href{https://containers.dev}{development containers}. Powered by \href{https://github.com/features/codespaces}{GitHub Codespaces} and \href{https://code.visualstudio.com/}{Visual Studio Code} (VS Code), these containers can run both locally and on the cloud, providing contributors with a consistent, ready-to-use workspace. 

The rest of this section offers a more detailed overview of some of PyPackIT's key components, features and capabilities.

\subsection{Continuous Configuration Automation}
\label{section-controlCenter}

PyPackIT streamlines project configuration, customization, and management through a centralized control mechanism based on DevOps practices like Infrastructure-as-Code (IaC). It provides a singular user interface to manage the entire project, and even multiple projects at once. PyPackIT's control center unifies all project configurations, metadata, and variables (Table \ref{table:control-center-options}) as declarative definitions in YAML—a standard human-readable data serialization format. These are consolidated into one repository location under version control, allowing for easy tracking of settings throughout the project's lifespan, and eliminating the need for manual settings and multiple configuration file formats and locations. Using Application Programming Interfaces (APIs) and dynamically generated files, PyPackIT automatically applies all settings to corresponding components, making the entire project dynamic and easily customizable. To further simplify project configuration and content management, the control center is equipped with several key features:

\begin{itemize}
    \item \textbf{Preconfiguration}: Default settings are provided based on current standards and best practices, requiring only project-specific metadata to be declared.
    \item \textbf{Augmentation}: Project information and statistics are generated at runtime by analyzing the repository and retrieving information from web APIs, minimizing manual inputs.
    \item \textbf{Templating}: Dynamic data generation and reuse is enabled for the entire repository, eliminating redundancy and allowing for centralized and automatic updates.
    \item \textbf{Inheritance}: Configurations can be inherited from online sources, ensuring consistent creation and centralized maintenance of multiple projects with shared settings.
    \item \textbf{Customization}: Additional settings and workflow routines can be added in YAML files or via Python plugins executed at runtime, maximizing customizability.
    \item \textbf{Validation}: Inputs are thoroughly verified against predefined schemas, providing comprehensive error reports for any inconsistencies.
    \item \textbf{Synchronization}: Changes are automatically propagated throughout the project, ensuring consistency without manual intervention.
    \item \textbf{Caching}: Intermediate data can be cached both locally and on GHA to reduce web API usage and speed-up processing, with configurable retention periods.
\end{itemize}

\begin{table*}[t]
\centering
\caption{Overview of PyPackIT's control center configuration options. The control center provides numerous options, enabling extensive customization.}
\label{table:control-center-options}
\begin{tabularx}{\textwidth}{>{\raggedright\arraybackslash}p{0.22\textwidth} >{\raggedright\arraybackslash}X}
\toprule
\rowcolor{white} \textbf{Category} & \textbf{Examples}\\
\midrule

Project descriptors & Name, title, abstract, keywords, and highlights.\\

Project metadata & License, citation, funding, team, roles, and contact information.\\

Package and test suite & Build configurations, package metadata, requirements, dependencies, and entry points.\\

Documentation & Dynamic READMEs and community health files, API references, changelogs, release notes, and other website configurations.\\

Issue tracking & GHI settings, issue and discussion forms, labels, pull request templates, and development documents.\\

Version control & Git/GitHub repository settings, branch/tag configurations and protection rules, dynamic files, and commit types.\\

Workflows & Continuous pipeline and tool settings, development environment specifications, and governance model.\\
\bottomrule
\end{tabularx}
\end{table*}

\subsection{Python Package}

PyPackIT provides a solid foundation for developing scientific Python applications aligned with the latest guidelines from the \href{https://www.pypa.io/}{Python Packaging Authority} (PyPA) \cite{PythonPackagingUserGuide}. It offers a build-ready \href{https://packaging.python.org/en/latest/glossary/#term-Import-Package}{import package} containing key source files with basic functionalities for data file support, error handling, and command-line interface development. Users simply need to extend these files with their application code, while PyPackIT's control center automatically provides all package configurations and allows for dynamic modification of modules, e.g., to add docstrings, comments, and code snippets. In addition to pure-Python packages, PyPackIT also supports packages with extension modules and non-Python dependencies. This is crucial for many scientific applications that rely on extensions and dependencies written in compiled languages like C and C++ for compute-heavy calculations. All necessary build configuration files such as \href{https://packaging.python.org/en/latest/specifications/pyproject-toml/}{\texttt{pyproject.toml}} and \href{https://docs.conda.io/projects/conda-build/en/latest/concepts/recipe.html}{Conda-build recipes} are automatically generated using control center settings, providing extensive project metadata to facilitate packaging, publication, and installation in PyPI and Anaconda ecosystems. Unique identifiers such as version numbers, DOIs, and release dates are also automatically generated for each build and added to the package. This dynamic package skeleton allows users to focus on writing scientific code, while PyPackIT ensures quality and FAIRness by automatically taking care of other tasks such as testing, refactoring, licensing, versioning, packaging, and distribution to multiple indexing repositories with comprehensive metadata and persistent identifiers.

\subsection{Test Suite}

PyPackIT simplifies software testing by providing a fully automated testing infrastructure built on \href{https://pytest.org/}{Pytest}, a widely-used Python testing framework. This setup supports unit, regression, end-to-end, and functional testing, which are integrated into PyPackIT's Continuous pipelines and automatically executed during software integration, deployment, and maintenance. For writing tests, PyPackIT provides a separate Python package skeleton next to the project's main package. This test suite includes the same automation and preconfiguration features as the main package, requiring users to only add test cases. It can be easily installed via Pip and Conda, replacing complex testing environment setups. PyPackIT also adds a command-line interface to the test suite, simplifying its execution by abstracting the details of running Pytest. During each release, the test suite is automatically licensed, versioned, packaged, and distributed along the software, allowing users to run tests and benchmarks on their machines and get detailed reports. Moreover, isolating the tests from the main package improves compatibility and reduces the overall size of the software. Structuring tests into a package also enhances modularity and reusability, simplifying the development of new test cases.

PyPackIT generates detailed test results and coverage reports in both machine and human-readable formats, using plugins such as \href{https://github.com/pytest-dev/pytest-cov}{Pytest-cov}. These reports are made available on the repository's GHA dashboard for each run, and can be automatically uploaded to coverage analysis platforms like \href{https://codecov.io/}{Codecov} to provide deeper insights into testing effectiveness. Testing outcomes are also automatically reflected in PRs and READMEs via dynamic badges and notifications, informing users about the research software's health status. By default, PyPackIT also enables branch and tag protection rules requiring all PRs to pass quality assurance and testing status checks before they can be merged into the mainline.

\subsection{Documentation}

\begin{figure*}[h!]
    \centering
    \includegraphics[width=1\linewidth]{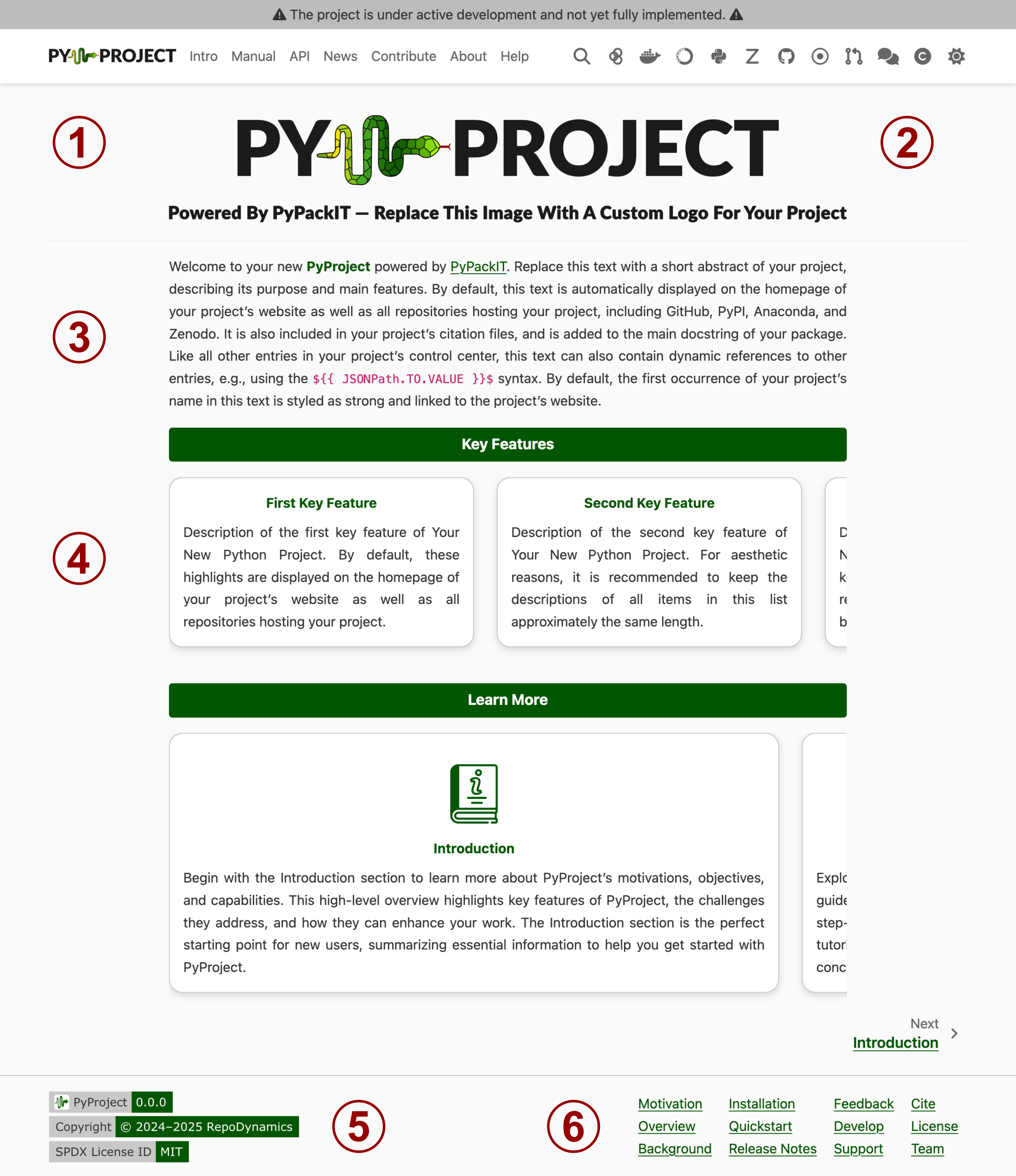}
    \caption{Default homepage of the project documentation website generated by PyPackIT. Circled numbers mark dynamic elements that are automatically updated according to control center configurations: 1) logo; 2) links to external resources on GitHub and other indexing repositories; 3) abstract; 4) highlights; 5) license and copyright; and 6) links to important website pages.}
    \label{fig:website}
    \vspace{2pt}
\end{figure*}

PyPackIT offers an automated solution for generating, deploying, and maintaining a comprehensive documentation website \cite{TenSimpleRulesForDocumentingSciSoft, WhatMakesCompSoftSuccessful}, significantly reducing the manual effort down to writing docstrings and brief software descriptions. It includes a fully configured \href{https://www.sphinx-doc.org}{Sphinx} website, designed according to a standardized layout for scientific software using the \href{https://pydata-sphinx-theme.readthedocs.io/}{PyData theme}. Website configurations and content are managed by PyPackIT's control center, allowing for easy customization and consistent updates. Moreover, \href{https://jinja.palletsprojects.com/}{Jinja} templating is enabled for all documentation files, with access to the entire control center. This enables the creation of sophisticated dynamic content from control center data, allowing PyPackIT to automatically generate project-specific documentation materials that are continuously updated throughout the development process. An overview is provided in Table \ref{table:dynamic-docs}, while Figure \ref{fig:website} shows the default website landing page as an example.

The website is further equipped with plugins such as \href{https://myst-nb.readthedocs.io/en/latest/}{MyST-NB}, \href{https://sphinx-design.readthedocs.io/}{Sphinx-Design}, and \href{https://sphinxcontrib-bibtex.readthedocs.io/}{SphinxContrib-BibTeX}, which enable executable code blocks, diagrams, charts, math notation, figures, tables, bibliographies, and responsive elements like admonitions, grids, and drop-down components. These are available via simple directives in powerful markup languages like \href{https://mystmd.org/}{MyST Markdown} and reStructuredText (reST), allowing for rich documentation of computational studies and research software. Moreover, a news blog with support for commenting, searching, content categorization and RSS feeds is integrated using \href{https://ablog.readthedocs.io/}{ABlog} and \href{https://giscus.app/}{Giscus}, establishing communication between the project and its community. PyPackIT can also automate blog posting for events like new releases and critical bug fixes, further simplifying the documentation effort.

\begin{table*}[t]
\centering
\caption{Overview of key documentation content dynamically generated and maintained by PyPackIT.}
\label{table:dynamic-docs}
\begin{tabularx}{\textwidth}{p{0.23\textwidth} X}
\toprule
\rowcolor{white} \textbf{Content Type} & \textbf{Details}\\
\midrule

Project information & License, copyright, citation, contributors, governance model, and contact information.\\

Package metadata & Python and OS requirements, dependencies, entry points, and versioning scheme.\\

Installation guide & Detailed instructions on how to install the application from available sources.\\

API reference & Documentation of all package components extracted from source code docstrings.\\

Changelogs & Chronological record of the entire development process, auto-generated from issue tickets, PR data, and configuration changes.\\

Release notes & Detailed summaries for each release providing a complete list of changes along with related identifiers and links to resources.\\

Contribution guide & Manuals on how to report issues, request changes, contribute artifacts, and maintain the software project. \\

Support guide & Instructions on how to find resources, fix common problems, ask questions, and seek further support.\\

\bottomrule
\end{tabularx}
\end{table*}

PyPackIT automates the build and deployment of the website through its CI/CD pipelines. Changes affecting the website trigger an automatic build process, which provides a preview for review. Upon approval, the updated website is deployed on \href{https://pages.github.com/}{GitHub Pages}—a free static web hosting service—accessible online through a default \texttt{github.io} domain. Users can also seamlessly specify a custom domain in the control center, if preferred. Additionally, hosting on the \href{https://readthedocs.org/}{Read The Docs} platform is supported with minimal setup, requiring only an account configuration. To improve project visibility and searchability, \href{https://ogp.me/}{Open Graph} metadata is embedded into each page using the \href{https://github.com/wpilibsuite/sphinxext-opengraph}{SphinxExt-OpenGraph} plugin, enabling search engine optimization (SEO) and rich social media previews.

Furthermore, PyPackIT can dynamically generate Markdown documents in various flavors, from declarative instructions in the control center. This is mainly used to add platform-specific \href{https://docs.github.com/en/communities/setting-up-your-project-for-healthy-contributions/creating-a-default-community-health-file#supported-file-types}{community health files} and \href{https://docs.github.com/en/repositories/managing-your-repositorys-settings-and-features/customizing-your-repository/about-readmes}{README files} for GitHub and other indexing repositories like PyPI. The READMEs serve as the front page of the repositories, featuring a clean design with key project information and links to the main sections of the documentation website. Moreover, they include a set of dynamic badges \cite{RepositoryBadges} that provide an up-to-date overview of project specifications, status, and progress.

\subsection{Version Control}

PyPackIT fully integrates with Git to automate tasks like branch and commit management, versioning, tagging, and merging. Motivated by well-established models such as GitLab Flow \cite{GitHubFlow, GitLabFlow}, PyPackIT adopts a branching strategy that supports the rapid evolution of research software in line with Agile and Continuous software engineering practices \cite{CICDSystematicReview}. Importantly, the model supports reproducibility and sustainability of computational studies relying on earlier releases, while enabling frequent experimentation and orthogonal development. It includes persistent release branches hosting stable versions, alongside transient branches for simultaneous development and deployment of multiple prereleases (Table \ref{table:branching-model}).

\begin{table*}[t]
\centering
\caption{PyPackIT's branching model. All branches are automatically generated as needed during the development process.}
\label{table:branching-model}
\begin{tabularx}{\textwidth}{p{0.15\textwidth} X}
\toprule
\rowcolor{white} \textbf{Branch Type} & \textbf{Description} \\
\midrule

Main & A special release branch that always contains the latest final version of the software, with its control center acting as the source of truth for all non-version-specific project configurations. \\

Development & Created from target release branches, providing an environment for integrating new changes before merging back to production. Separate development branches allow simultaneous development of multiple features and release candidates. \\

Prerelease & Created from development branches for publishing prerelease versions, facilitating updates and documentation during the prerelease period before final merging into a release branch. \\

Release & Created from the main branch before merging backward-incompatible changes, to preserve earlier major versions for long-term maintenance and support while enabling rapid project evolution. \\

\bottomrule
\end{tabularx}
\end{table*}

\begin{figure}[h]
    \centering
    \includegraphics[width=0.8\linewidth]{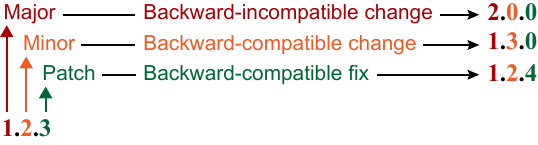}
    \caption{Schematic overview of Semantic Versioning. 
    Version numbers follow the \texttt{X.Y.Z} format, where \texttt{X}, \texttt{Y}, and \texttt{Z} represent major, minor, and patch numbers, respectively. The major number is incremented for backward-incompatible changes, the minor number for backward-compatible changes, and the patch number for bug fixes. For each release, one of these components is incremented by 1, while the ones to its right are reset to 0. The public API is introduced in version \texttt{1.0.0}, while major version zero (\texttt{0.y.z}) is for initial development and signals an unstable API.}
    \label{fig:semver}
\end{figure}

Moreover, to provide scientific software with a clear and transparent development history and ensure its long-term findability and accessibility, all developmental, pre-, and final releases are automatically versioned and tagged. For this, PyPackIT implements a versioning scheme based on \href{https://semver.org/}{Semantic Versioning} (SemVer), which clearly communicates changes to the library's public API (Figure \ref{fig:semver}) \cite{EmpComparisonOfDepIssues, WhatDoPackageDepsTellUsAboutSemVer}.

\begin{figure}[t]
    \centering
    \includegraphics[width=1\linewidth]{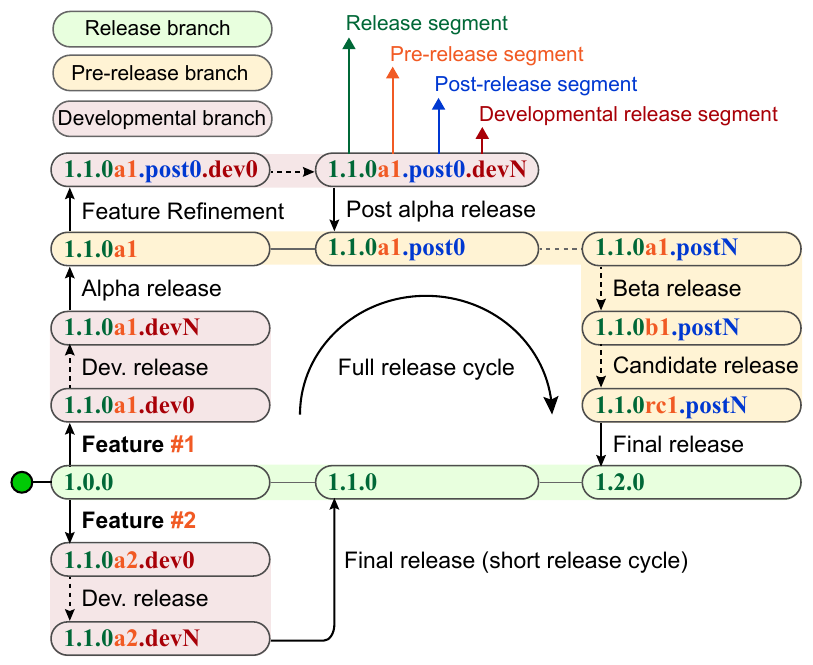}
    \caption{PyPackIT's version control strategy is demonstrated with an example starting from final release version \texttt{1.0.0}. Two individual features (issue ticket numbers 1 and 2) are simultaneously implemented in separate development branches, where each iteration is published as a developmental release with a unique version number (green: release segment, orange: prerelease segment, and red: developmental release segment). Note the incorporation of issue ticket numbers into the prerelease segments. Feature 2—demonstrating a short release cycle—is merged directly into the release branch as a final release, while Feature 1 undergoes a prerelease phase with with further refinements published as post-releases. After alpha, beta, and release candidate stages, it is merged back into the original release branch as a final release. The final version is automatically redetermined during merging so that \texttt{1.1.0rc1.postN} is released as final version \texttt{1.2.0}, since \texttt{1.1.0a2.devN} was already released as \texttt{1.1.0}.}
    \label{fig:versioning}
\end{figure}

PyPackIT uses correlated issue ticket data to determine whether changes correspond to a new major, minor, or patch release. It then automatically calculates the full version number by comparing the latest version tags in the development branch and its target release branch. Subsequently, for automatic deployment on indexing repositories such as PyPI and Anaconda, PyPackIT generates a canonical public version identifier in accordance with the \href{https://packaging.python.org/en/latest/specifications/version-specifiers/}{latest PyPA specifications} (Figure \ref{fig:versioning}). This is done by adding additional segments to the SemVer release number in order to uniquely identify all developmental, pre-, and post-releases. By incorporating the correlated issue ticket number as a unique identifier in prerelease versions, PyPackIT enables scientific projects to simultaneously develop and publish multiple release candidates, for example to test different approaches for solving the same problem. Publishing developmental and prereleases allows the community to thoroughly investigate upcoming changes and provide feedback, improving the quality of scientific software and reducing the risk of using buggy applications in research \cite{CICDSystematicReview}.

\subsection{Issue Management}

PyPackIT establishes an automated pull-based development workflow
based on a well-tested strategy for scientific projects \cite{ConfigManageForLargescaleSciComp}: New tasks in the project start by submitting a ticket to its ITS, promoting community collaboration while ensuring thorough documentation of software evolution. PyPackIT automatically configures and maintains GHI according to customizable configurations in the control center. By default, it includes a comprehensive set of issue forms allowing users to submit type-specific tickets, such as bug reports and feature requests for the application or its test suite and documentation. Designed according to best practices \cite{WhatMakesAGoodBugReport, NeedsInBugReports, QualityOfBugReportsInEclipse}, these forms collect essential type-specific user inputs in a machine-readable format. They also include selectable options for inputs such as currently available version numbers and API endpoints, which are automatically kept up to date by PyPackIT's CCA pipeline.

After an issue is submitted, PyPackIT processes user inputs to automate the bulk of issue management activities. It uses a customizable template to generate a standardized software development protocol for the issue, which is attached to the ticket. By default, this document includes formatted user inputs under User Requirements Document (URD), and contains other standard sections like Software Requirements Document (SRD) and Software Design Document (SDD) for maintainers' inputs. It also includes dynamic components that are updated by PyPackIT to reflect current status and progress, such as a list of completed/remaining tasks, related pull requests, and activity logs. To improve ITS organization and searchability, PyPackIT uses inputs to automatically label tickets based on issue type, affected versions, API endpoints, and other configurable options. Ticket inputs can also be used to automatically assign specific team members to each ticket, as defined by the project's governance model in the control center.

After triage, project members can communicate with PyPackIT through issue comments and labels. For example, by posting a semantic comment under the issue, members can command PyPackIT to automatically test a certain version of the software in a specific environment with given test cases. This greatly accelerates bug triage, eliminating the need for manual branch creation, test suite modification, and execution. If the new test cases fail, PyPackIT automatically adds them to the test suite and initiates a bug fix cycle, streamlining the process of turning bugs into tests to validate fixes and prevent future recurrences \cite{BestPracticesForSciComp, 10SimpleRulesOnWritingCleanAndReliableSciSoft}. Furthermore, changing the ticket's status label tells PyPackIT to perform additional tasks. Rejected tickets are automatically closed with updated documentation, while tickets labeled ready for implementation trigger the creation of development branches from affected release branches. These branches are automatically prepared for development and linked to the issue ticket on GitHub. For example, a new entry is added to the changelog file, and the README file is temporarily modified to reflect development on the branch. Changes are added with a commit message containing issue data, ensuring a clear and informative history on Git.

\subsection{Continuous Integration} 
\label{overview-ci}

When an issue is ready for implementation, PyPackIT opens a draft PR from each created development branch. Similar to issue tickets, PRs are also automatically labeled, assigned, and documented. If the PR corresponds to a new release, PyPackIT also creates draft releases on selected platforms like GitHub Releases and Zenodo. Contributors can thus immediately start the implementation in the provided development containers, requiring only a web browser. With each commit on a development branch, PyPackIT runs its CI pipeline to integrate changes into the codebase according to best practices \cite{CICDSystematicReview, ModelingCI, ContinuousSoftEng}, while generating comprehensive build artifacts and reports for review. The tasks performed by PyPackIT's CI pipeline are summarized in Table \ref{table:ci-tasks}. 

\begin{table*}[h!]
\centering
\caption{Overview of common integration tasks automated by PyPackIT's CI pipeline. Similar to other project components, the pipeline is highly customizable through PyPackIT's control center, allowing users to modify defaults or add additional tasks.}
\label{table:ci-tasks}
\begin{tabularx}{\textwidth}{p{0.13\textwidth} X}
\toprule
\rowcolor{white} \textbf{CI Task} & \textbf{Description} \\
\midrule
CCA & Dynamic files and content in the branch are synchronized with updated control center configurations.\\

Formatting & Changed source files are formatted according to Python's official style guide \cite{PEP8}, using \href{https://docs.astral.sh/ruff/}{Ruff} as a fast drop-in replacement for the popular \href{https://black.readthedocs.io/}{Black} formatter.\\

Code\newline analysis & Static code analysis and type checking is performed to detect code violations, errors, and security issues, using well-established linters like \href{https://docs.astral.sh/ruff/}{Ruff}, \href{https://mypy.readthedocs.io/}{Mypy}, and \href{https://codeql.github.com/}{CodeQL}. \\

Data\newline validation & Data files in JSON, YAML, and TOML formats are checked for issues and syntax errors. \\

Refactoring & Available fixes such as end-of-file and end-of-line standardization and safe refactoring suggestions by Ruff are automatically applied to affected files.\\

Dependency\newline review & Changed dependencies are analyzed for security and license issues, using GitHub's \href{https://github.com/actions/dependency-review-action}{Dependency-Review action}. \\

Build & Source (\href{https://packaging.python.org/en/latest/glossary/\#term-Source-Distribution-or-sdist}{sdist}) and built (\href{https://packaging.python.org/en/latest/glossary/\#term-Built-Distribution}{wheel}) distributions are generated for package and test suite according to \href{https://packaging.python.org/en/latest/tutorials/packaging-projects/\#generating-distribution-archives}{PyPA guidelines}. PyPackIT uses PyPA's \href{https://build.pypa.io/}{Build} and \href{https://cibuildwheel.pypa.io/}{Cibuildwheel} tools to create platform-independent wheels for pure-Python packages, and platform-dependent binaries for packages with extension modules. Conda distributions are also generated using the \href{https://docs.conda.io/projects/conda-build/en/stable/}{Conda-build} tool.\\

Containerization & A \href{https://jupyter.org/}{Jupyter}-enabled Docker image is created using \href{https://jupyter.org/hub}{JupyerHub}'s \href{https://github.com/jupyterhub/repo2docker}{Repo2docker} application, with the package, test suite, and dependencies installed. Besides Python and Conda, Debian packages can be installed from \href{https://wiki.debian.org/apt-get}{APT}, enabling easy distribution of applications with complex environments.\\

Testing & Test suite is executed on all supported operating systems and Python versions to verify the correctness and compatibility of applied changes. This is done for both Python and Conda distributions, as well as the Docker image. \\

Website \newline build & Documentation website is built with the latest changes and attached to the CI for offline preview. If deployment to Read The Docs is enabled, a link for online preview is also added to the PR.\\

Changelog \newline update & Project's changelog file is updated with data from the corresponding issue, PR, and commits, along with identifiers such as DOI, version, commit hash, and date. The changelog contains machine-readable data in JSON, maintaining a chronological record of the entire project evolution. It is used to automatically generate release notes in Markdown format, according to a customizable template.  \\

Draft\newline update & All draft releases are updated with the latest configurations, documentation, and build artifacts. Release-specific metadata such as external contributors and acknowledgments can also be added using automated rules or semantic PR comments. \\

Progress \newline tracking & Dynamic elements in the issue ticket and draft PR are updated to reflect the progress. PyPackIT uses commit messages to automatically mark corresponding tasks in the PR as complete. As GHI displays the number of completed and remaining tasks in each issue ticket/PR, this provides a clear overview of overall progress in the project.\\

Report & Comprehensive logs and reports are generated for each step, improving the visibility of integration status and facilitating reviews. These are displayed on GHA's terminal in real-time, and rendered as a responsive HTML document for download.\\

\bottomrule
\end{tabularx}
\end{table*}

Furthermore, when release-related changes are successfully integrated, a new developmental release is automatically versioned, tagged, and published to selected indexing repositories, e.g., Anaconda, TestPyPI, Zenodo Sandbox, and Docker registries. This allows the application to be automatically tested in a production-like environment, ensuring that it works as intended after download and installation on user machines. It also greatly simplifies the sharing of new developments among collaborators, enabling feedback and reviews during the implementation phase. Moreover, the entire development progress leading to each final release is clearly documented and permanently available \cite{CICDSystematicReview}.

When all implementation tasks specified in the PR are marked as complete, PyPackIT automatically initiates the review process by sending review requests to designated members according to the project's governance model. The CI pipeline is automatically triggered by changes during the review process as well, providing team members with detailed status checks, reports, and build artifacts to keep them informed about the outcomes of revisions. Upon PR approval by reviewers, PyPackIT automatically merges the changes into the base branch. By default, this is done by squash merging the development branch into the (pre)release branch to maintain a clear and linear Git history. The commit message is automatically generated according to a template, and can include issue ticket number and other details to establish issue–commit links and reflect documentation in the VCS. Moreover, when changes are merged into the main branch, all project-wide configurations such as repository and workflow settings are updated according to control center settings.

\subsection{Continuous Deployment} \label{overview-cd}

PyPackIT's CD pipeline is activated when the merged changes correspond to a new (pre)release. It generates a public version number to tag the release and deploy it to user-specified indexing repositories. By default, the built Docker image is published to GHCR, providing an isolated environment for running the application on any machine or HPC cluster. This greatly facilitates the reproducibility of computational studies, allowing scientific applications to always run in the exact same software environment they were developed for. PyPackIT also uses the Docker image to trigger a build on \href{https://mybinder.org}{mybinder.org}—a free \href{https://binderhub.readthedocs.io}{BinderHub} instance allowing users to interact with the image from their web browser using a JupyterLab interface. The build is cached on \texttt{mybinder.org} servers, significantly shortening subsequent loading times.

Further deployment tasks include finalizing and publishing the draft releases on Zenodo and GitHub Releases, which can contain any number of user-specified source and build artifacts. By default, they include all distribution packages, built documentation, citation and license files, and a \texttt{Dockerfile} pointing to the deployed Docker image. A DOI is reserved with Zenodo prior to deployment and subsequently included in the metadata of all distributions. Each release is therefore permanently available through a unique DOI, enabling reliable citations. FAIRness is further improved by uploading source and built distributions to PyPI and Anaconda repositories. This allows for customized and reproducible builds while facilitating installation on different platforms and architectures using Pip and Conda package managers. Releases are accompanied by up-to-date metadata, documentation, and detailed release notes, facilitating discovery and accessibility through various keywords, classifiers, and identifiers. The updated documentation website is deployed online, with added banners and blog posts to announce the new release.

\subsection{Continuous Maintenance, Refactoring, and Testing}

In addition to CI/CD pipelines that integrate and deploy new changes, PyPackIT also periodically runs other Continuous pipelines to ensure the long-term sustainability of the project and its products. For example, previous releases are tested with the latest versions of their dependencies to confirm compatibility with up-to-date environments. The CCA pipeline is run to update dynamic configurations and content, such as those inherited from external sources. Code analysis, refactoring and style formatting tasks are performed as well, using updated tools and standards to curb the ever-increasing code complexity and maintain quality and consistency during both development and support phases. To maintain the health of the development environment, the repository and its components are frequently cleaned up, removing outdated artifacts, such as builds, logs, and reports.

During each run, PyPackIT automatically applies updates and fixes to a new branch, and creates a PR for review by maintainers. After approval, changes are merged into the project and applied to all components. Alternatively, users can opt in for automatic merging without the need for manual approvals. Similarly, if a problem is found, a new issue ticket is submitted to the issue tracking system, notifying project maintainers to take action. These automated processes simplify and encourage maintenance activities \cite{CanAutoPRsEncourageDepUpgrade}, facilitating the prolonged development and support of research software.

\subsection{Licensing}

PyPackIT streamlines project licensing and copyright management according to the best practices for research software \cite{QuickGuideToLicensing, BarelySufficientPracticesInSciComp, SustainableResearchSoftwareHandOver, ShiningLight, 10RuleForSoftwareInCompBio}. In line with the upcoming Python Enhancement Proposal (PEP 639) for improving license clarity \cite{PEP639}, PyPackIT adopts the System Package Data Exchange (\href{https://spdx.org/}{SPDX}) license standard. This allows users to define complex licenses using simple \href{https://spdx.github.io/spdx-spec/v3.0.1/annexes/spdx-license-expressions/}{SPDX license expressions}. PyPackIT supports all \href{https://spdx.org/licenses/}{SPDX License List} entries as well as user-defined licenses, and can customize licenses with project-specific information.

By default, new projects are licensed under the \href{https://spdx.org/licenses/MIT.html}{MIT License}—a permissive \href{https://opensource.org/licenses}{OSI-approved} and \href{https://www.gnu.org/licenses/license-list.en.html}{FSF Free} license supporting research software commercialization \cite{SettingUpShop} and fulfilling the Bayh–Dole requirements for patenting publicly funded products \cite{BayhDole}. Another recommended option is the GNU Affero General Public License (\href{https://www.gnu.org/licenses/agpl-3.0.en.html}{AGPL-3.0}) that promotes open science by enforcing downstream source disclosure. To change the project license, users only need to provide the corresponding expression in the control center, while PyPackIT automatically downloads and integrates the required license data, as summarized in Table \ref{tab:licensing}.

\begin{table*}[t]
    \centering
    \caption{PyPackIT's automated license and copyright management workflow.}
    \label{tab:licensing}
    \begin{tabularx}{\textwidth}{p{2.4cm} X}
        \toprule
        \rowcolor{white} \textbf{Task} & \textbf{Description} \\
        \midrule

        Validation & User is notified if the provided license expression is invalid, uses deprecated or obsoleted components, or has conflicts with project dependencies. \\
        
        Customization & Placeholder values in license content, such as project name, copyright notice, and contact information are automatically replaced with project information according to control center configurations. \\
  
        Documentation & For each license component, syntactically valid and nicely formatted documents are generated and added to the repository according to \href{https://github.blog/changelog/2022-05-26-easily-discover-and-navigate-to-multiple-licenses-in-repositories/}{GitHub specifications} to be correctly recognized and displayed. These are also included in the project's website, along with other license details. \\
        
        Annotation & An SPDX \href{https://spdx.dev/learn/handling-license-info/}{short-form identifier} is added as a comment in all source files and as a footer badge in all documentation files, clearly communicating license information in a standardized human and machine-readable manner. A short copyright notice can also be automatically added to all or certain module docstrings. \\
       
        Distribution & License files are automatically included in all releases and distributed with the package and test suite. Platform-specific license metadata and identifiers are also added to facilitate license identification by indexing services and package managers.\\

        \bottomrule
    \end{tabularx}
\end{table*}

\subsection{Security}

To enhance project security while supporting community collaboration, PyPackIT incorporates several security measures, outlined in Table \ref{tab:security-measures}. Moreover, to ensure that PyPackIT itself is secure, its entire infrastructure is natively implemented and self-contained. With the exception of well-established GHA actions and Python libraries from trusted vendors like GitHub and PyPA, PyPackIT does not rely on other third-party dependencies. This allows the PyPackIT development team to rapidly respond to issues and continuously improve PyPackIT, while making it fully transparent and easily auditable by the community.
\begin{table*}[t]
    \centering
    \caption{Overview of PyPackIT's security measures.}
    \label{tab:security-measures}
    \begin{tabularx}{\textwidth}{>{\raggedright\arraybackslash}p{2.4cm} X}
        \toprule
        \rowcolor{white} \textbf{Security Measure} & \textbf{Description} \\
        \midrule
        
        Protection rules & Release branches and version tags require passing status checks and admin approvals before changes can be made, ensuring repository integrity. \\

        Vulnerability scanning & Code analysis tools integrated into CI/CD pipelines ensure changes do not introduce security vulnerabilities. \\

        Dependency monitoring & Project dependencies are continuously monitored, alerting maintainers of issues and proposing updates. \\
      
        PR approvals & External PRs require project maintainers' approval before CI/CD pipelines can run, preventing the execution of untrusted code. \\
   
        Workflow security & Workflows are developed according to best practices \cite{AutoSecurityAssessOfGHAWorkflows, GHADocsSecurity} to prevent security issues like command injection, token/secret exposure, and untrusted applications. \\
        
        \bottomrule
    \end{tabularx}
\end{table*}

\subsection{Publication}

PyPackIT also facilitates scientific publications, enabling users to generate papers, posters, and presentations in a variety of formats including LaTeX and Markdown. To achieve this, PyPackIT provides a dedicated development container that includes a full installation of \href{https://tug.org/texlive/}{Tex Live} and \href{https://pandoc.org}{Pandoc}, allowing for seamless conversion and compilation of documents to PDF and other output formats. 
This setup enables researchers to write and compile documents in a manner similar to online TeX-editing platforms like \href{https://www.overleaf.com/}{Overleaf}, but with the added benefit of being integrated with their codebase. 
The container includes powerful tools and extensions like \href{https://github.com/James-Yu/LaTeX-Workshop}{LaTeX-Workshop}, which enable continuous compilation and preview in GitHub Codespaces and VS Code. This allows researchers to maintain their publications alongside their code, leveraging features like version control, detailed change histories and comprehensive revisions through commit messages, issue tickets, and pull request reviews. Furthermore, PyPackIT supports automation, enabling the dynamic generation of figures and tables directly from code, ensuring consistency between the software and its related publications.

\section{Summary}
\label{section-summary}

We developed PyPackIT: an open-source, ready-to-use, cloud-based automation tool that simplifies every step of the research software development life cycle, from initiation and configuration to publication and long-term maintenance. PyPackIT ensures that scientific software projects are FAIR (Findable, Accessible, Interoperable, Reusable) and aligned with the latest software engineering best practices. To facilitate interoperability, reusability, and adoption, PyPackIT specializes in the production of research software in Python, the leading programming language for scientific applications. Promoting accessibility and collaboration, PyPackIT supports open-source development on GitHub. PyPackIT is readily installable in both new and existing GitHub repositories, providing them with a comprehensive and robust project infrastructure that includes:

\begin{itemize}
    \item \textbf{Control Center}: A centralized user interface to automatically manage all project metadata and settings according to DevOps practices. It enables dynamic content management for the entire project, simplifying setup, configuration, and maintenance via built-in templating, inheritance, and synchronization features that apply modifications to all project components. Preconfigured according to research software requirements, it only requires users to fill few project-specific metadata, while allowing for extensive customization.
    \item \textbf{Python Package}: A customized build-ready package skeleton with all necessary source and configuration files. Users only need to add scientific code to the provided source files, while PyPackIT automates all packaging, versioning, licensing, build, containerization, distribution, and indexing tasks. To facilitate research software findability and usage, PyPackIT supports automatic deployment to various indexing repositories, including Anaconda, PyPI, Zenodo, GitHub Releases, and Docker registries. Using Zenodo, a unique DOI is registered for each release, facilitating citation and reproducibility of computational studies.
    \item \textbf{Test Suite}: A testing infrastructure enables immediate adoption of test-driven methodologies. Requiring users to only write test cases in the provided package, PyPackIT automatically performs unit, regression, end-to-end, and functional testing throughout the development life cycle. It ensures research software quality and correctness, while improving awareness of software health status via comprehensive reports. The test suite is automatically packaged and distributed along each release, facilitating the reproducibility of test results for verification of software functionality and performance.
    \item \textbf{Documentation}: A documentation website is deployed on GitHub Pages and Read The Docs platforms, leaving users with minimal documentation tasks such as writing issue tickets and docstrings. The website is filled with customized documentation including project and package information, installation guide, API reference, changelogs, release notes, and other developer and user manuals. Most website content are dynamic and easily customizable via PyPackIT's control center. PyPackIT also enables the generation of dynamic LaTeX and Markdown documents, such as scientific papers or repository README and community health files.
\end{itemize}

After installation, PyPackIT automatically activates in response to various repository events, executing appropriate tasks on the GitHub Actions cloud computing platform. It establishes a pull-based software development workflow for collaborative and highly iterative development of research software. PyPackIT's workflow includes comprehensive Continuous software engineering pipelines that use up-to-date tools and technologies to enable a cloud-native Agile software development process. They automate the bulk of repetitive engineering and management activities throughout the software life cycle, including:

\begin{itemize}
    \item \textbf{Issue Management}:  The project's issue tracking system is dynamically maintained, supplied with specialized submission forms to collect type-specific user inputs in a machine-readable format. This allows PyPackIT to automate issue management activities such as ticket labeling and organization, bug triage, task assignment, development documentation, issue–\allowbreak commit linkage, and progress monitoring.
    \item \textbf{Version Control}: To enable rapid project evolution while ensuring the reproducibility of earlier results, PyPackIT implements a specialized branching model and versioning scheme for simultaneous publication and support of multiple releases. Fully integrating with Git, PyPackIT's workflow automates version control tasks such as branch and pull request creation, tagging, merging, and commit management.
    \item \textbf{Continuous Integration and Deployment}: PyPackIT's CI/CD pipelines automate tasks such as code style formatting, static code analysis, type checking, testing, build, containerization, and release, thus eliminating the need for dedicated integration and deployment teams, while increasing control, integrity, scalability, security, and transparency of the Agile development process.
    \item \textbf{Continuous Maintenance, Refactoring, and Testing}: PyPackIT periodically performs maintenance and monitoring tasks to ensure the long-term sustainability of projects and the health of development environments. Tasks include testing previous releases in up-to-date environments, updating dependencies and development tools, refactoring code, and cleaning up the repository to remove outdated artifacts such as builds, logs, and reports.

\end{itemize}

In summary, PyPackIT provides a comprehensive, dynamic, and highly customizable project skeleton for scientific Python applications along with a ready-to-use cloud-native development environment. This eliminates the need for manual project setup and configuration, enabling researchers to immediately begin the actual implementation of software, even directly from the web browser. PyPackIT's automated development workflow greatly simplifies the research software development process, reducing manual tasks to writing issue tickets, scientific code, test cases, and minimal documentation. It carries out crucial and repetitive software engineering activities on the cloud and consistently enforces best practices throughout the software life cycle, thereby improving product quality while reducing cost and effort. PyPackIT enables developers to focus solely on the scientific aspects of their project, diminishing common challenges such as limited funding, time, and technical expertise. As the scientific inquiry process increasingly relies on research software, PyPackIT can be a valuable asset to computational studies that are now integral to many fields.

\section{Author Contributions}

\textbf{AA}: Conceptualization, Investigation, Methodology, Software, Validation, Visualization, Writing - Original draft, Writing - Review and Editing. 
\textbf{RLR}: Investigation, Project administration, Supervision, Validation, Writing - Review and Editing.
\textbf{AV}: Conceptualization, Funding acquisition, Investigation, Project administration, Supervision, Validation, Writing - Review and Editing.

\section{Acknowledgments}

We are thankful for funding from Saarland University for the NEDD project. 
AV acknowledges financial support from a BIH Einstein Visiting Professor Fellowship, from where RLR's position is funded.

\bibliographystyle{elsarticle-num}
\bibliography{references.bib}
\end{document}